\documentclass[doublecol]{epl2} 

\title{Discreteness as ontology: A hodon-based approach to dark matter}
\shorttitle{Hodons as DM} 

\author{Arkady Bolotin}

\institute{                    
  Ben-Gurion University of the Negev, Beersheba (Israel)\\
}

\abstract{
This work proposes a geometric-statistical reinterpretation of the dark sector, grounded in a discrete spacetime framework composed of non-material spatial units termed hodons. Unlike particle-based dark matter models, hodons are kinematically inert and possess ultra-light effective mass derived from vacuum energy density and holographic volume bounds. We introduce a covariant scalar field $\mathcal{N}(x^\mu)$ representing local hodon density and derive an entropy-driven evolution equation consistent with causal structure and general relativity. The resulting stress-energy contribution from hodon fluctuations yields gravitational clumpiness without invoking new particles or modified gravity. A virial-based toy model demonstrates that baryonic matter surrounded by hodons forms stable, cored halo profiles, consistent with galactic rotation curves and low-mass halo observations. The framework naturally suppresses small-scale structure via spatial uncertainty relations, aligning with constraints from the Lyman-$\alpha$ forest and weak lensing. By integrating Bousso's covariant entropy bound and distinguishing between strong and weak holography, we situate the model within a broader epistemological context. These results suggest that dark sector phenomenology may emerge from the statistical geometry of space itself, offering a falsifiable alternative to particle dark matter.}

\begin{document}

\maketitle

\section{Introduction}

Dark matter (DM) is a hypothetical form of matter that does not interact directly with ordinary baryonic matter or radiation. Its existence is inferred from gravitational phenomena, including galaxy rotation curves and the large-scale structure of the universe -- phenomena that cannot be fully explained by general relativity without invoking a substantial unseen mass component \cite{Bertone, Read, Martino, Cacciatori}. Leading candidates for DM include unknown nonbaryonic particles -- both massive and light \cite{Gondolo} -- as well as primordial black holes \cite{Chen}.

While these strategies yield robust phenomenology, they often entail significant ontological inflation. This paper proposes an alternative path: the \textbf{hodon framework}, a speculative but principled model that treats dark matter as a structural artifact of discrete spacetime rather than a material entity.

Rooted in the premise that spacetime emerges from an informational and geometric substrate governed by Planck-scale discreteness, the framework offers conceptual economy by reframing gravitational clumpiness as a statistical manifestation of spatial uncertainty. Rather than postulating additional fields or exotic matter content, the hodon approach builds upon accepted ideas -- finite information density, holographic scaling, and observer complementarity -- to derive the observed gravitational effects from first principles.

While speculative, the framework serves a heuristic role: it generates testable predictions about vacuum energy, halo distribution, and spatial resolution bounds, consistent with observed cosmological regularities. Its speculative nature is not a departure from rigor, but a methodological necessity in exploring foundations where traditional tools are epistemically constrained. In this respect, it complements efforts in quantum gravity and discrete geometry while inviting a reevaluation of what constitutes explanatory sufficiency at fundamental scales.

This paper develops a minimal toy model to formalize these ideas, examines its consistency with empirical data, and positions the framework in dialogue with other foundational proposals. By doing so, it illustrates how speculative thinking -- when disciplined by conceptual clarity and mathematical rigor -- can be a productive engine for theoretical advancement.

\section{Finite geometry}

Several motivations for finitism and discreteness in geometry can be put forward.

One motivation stems from quantum theory, which introduces the hypothesis of quantization -- the idea that the magnitude of a physical property can take on only discrete values -- implying that space itself may possess a discrete structure \cite{Rovelli}. Another comes from the possibility that the discreteness of space could serve as a natural ultraviolet cutoff, effectively regularizing divergences in physical theories \cite{Malament}.

Furthermore, in accordance with the black hole entropy bound, only a finite number of degrees of freedom can be localized within a finite region of space \cite{Cao, Bolotin24}. Since quantum field theory (QFT) intrinsically links degrees of freedom to the fundamental elements of space -- namely, points -- this implies that the number of points within any finite-volume region must also be finite.

Motivated by these considerations, we propose that physical space is fundamentally discrete and can be described by \textbf{finite geometry}, a specific type of discrete geometry characterized by a finite set of points. Such points, often referred to as \textbf{hodons}, are typically described as the ``atoms of space'' \cite{Crouse1, Crouse2, Smolin}.

\subsection{Identifying hodons}

To address the problem of identifying hodons, Wheeler's ``It from Bit'' concept \cite{Wheeler} offers a compelling perspective. This concept suggests that the fundamental building blocks of reality are \textbf{informational} rather than material. In line with this view, each hodon -- as a fundamental unit of physical space -- can be identified with a single bit of information and, since one bit corresponds to a system with a single degree of freedom, be directly associated with a \textbf{single degree of freedom}.

Another argument supporting the identification of hodons as informational entities is that information -- unlike other physical properties -- is \textbf{naturally discrete}. Its magnitude is typically measured in bits, which are fundamentally discrete units. Therefore, associating hodons with information aligns logically with their inherently discrete nature.

Furthermore, in QFT, each mode $k$ of a quantized field behaves like a quantum harmonic oscillator, contributing a zero-point energy of $\frac{1}{2} \hbar \omega_k$, where $\omega_k$ is the angular frequency associated with that mode. Summed over all modes, these contributions give rise to pervasive vacuum fluctuations throughout empty space. This suggests that, in addition to encoding a unit of information, each hodon must also be intrinsically associated with vacuum energy. If $\omega = \sum_k \omega_k$ denotes the cumulative angular frequency, the corresponding vacuum energy per hodon may be represented as $E_{\mathrm{hodon}} = \frac{1}{2} \hbar \omega$.

In other words, since vacuum energy is a fundamental and ever-present attribute of space, each hodon -- as the ``atom of space'' - not only carries a discrete degree of freedom, but also embodies the lowest possible value (a quantum) of vacuum energy, $E_{\mathrm{hodon}}$.

\section{Discrete-to-continuous transition}

That said, any discrete geometry proposal must confront the challenge of how the continuum can re-emerge, even if only approximately. To formalize the transition from discrete hodon-based space to continuous geometrical observables, we proceed in four steps: \textbf{embedding}, \textbf{addressing anisotropy}, \textbf{averaging}, and \textbf{limiting}.

\subsection{Embedding hodons in classical regions}

Let $\mathcal{R}$ be a bounded region in a geometrical space. Such a space is a mathematical structure used in the theoretical and axiomatized studies of physical space (e.g., in the geometric theory of gravitation such as general relativity). A geometrical space is infinite, continuous, and defined by fundamental elements, i.e., points, that have no measurable properties.

Due to its nonmaterial nature, a hodon does not require a definite shape when represented (i.e., embedded) in a geometrical space. As a result, inquiries regarding the existence of distinct parts within a hodon are not applicable.

This implies that, within a geometrical space, the presence of exactly $n$ hodons in a given region can only be determined \textbf{probabilistically}. Consequently, one can evaluate only the likelihood that a specific region of geometrical space is associated with $n$ hodons. In other words, the number of hodons (degrees of freedom) mapped into (assigned to) a bounded region of geometrical space -- henceforth referred to as a ``classical region'' -- is represented as a random variable $N=n$.

\subsection{Addressing anisotropy}

To address {the problem of anisotropy in discrete geometry}, the mapping between discrete physical space and its corresponding geometrical space can be modelled using a \textbf{Poisson point process} (PPP). This approach envisions the distribution of hodons as a completely random sprinkling of discrete elements -- analogous to ink droplets randomly scattered on a surface \cite{Bombelli} -- across a classical region.

Formally, for a classical region $\mathcal{R}$ or its boundary $\partial\mathcal{R}$ with a measure $\mathcal{M}$ (interpreted as volume $V$ or area $A$, depending on context), the probability of observing exactly $n$ hodons is given by:

\begin{equation} \label{SPRIN} 
   P \!\left\{ N = n \right\}
   =
   \frac{ \Lambda^n \cdot e^{-\Lambda} } {n!}
   \;\;\;\;  ,
\end{equation}

where $\Lambda$ is the intensity parameter of the PPP, representing the expected number of hodons in the region:

\begin{equation} \label{LAMB} 
  \Lambda
   =
   \mathrm{E} \left[ N(\mathcal{R}) \right]
   =
   \int_{\mathcal{R}} \varrho(x) \,\upd \mathcal{M}(x)
   \;\;\;\;   .
\end{equation}

Here, $\varrho(x)$ is the local sprinkling density, and $\upd\mathcal{M}(x)$ denotes the infinitesimal measure element (either $\upd V$ or $\upd A$) over the region. In the case of a homogeneous PPP, the density $\varrho(x)$ is constant, and Eq.\,(\ref{LAMB}) simplifies to $\mathrm{E} \left[ N(\mathcal{R}) \right] = \varrho_{\mathcal{M}}\cdot\mathcal{M}$, where $\varrho_{\mathcal{M}}$ is the uniform sprinkling rate and $\mathcal{M}$ represents the total measure of the region $\mathcal{R}$.

\subsection{Hodon counts in classical regions}

Consider a classical region $\mathcal{R}\subset\mathrm{\bf{R}}^3$ of characteristic length $\ell$, volume $V\!\left(\mathcal{R}\right)\sim\ell^3$, and boundary surface area $A\mkern-0.75mu\left(\partial\mathcal{R}\right)\sim\ell^2$.

We assume that the expected number of hodons embedded in $\mathcal{R}$ is bounded by the holographic principle:

\begin{equation} \label{BOUND} 
   \mathrm{E} \left[ N\!\left(\mathcal{R}\right) \right]
   \le
   \frac{1}{4\ell_{\mathrm{P}}^2}
   \cdot
   A\mkern-0.75mu\left(\partial\mathcal{R}\right)
   \sim
   \frac{\ell^2}{4\ell_{\mathrm{P}}^2}
   \;\;\;\;  .
\end{equation}

\subsection{Statistical averaging and spatial uncertainty}

The holographic constraint allows us to define the effective volume per hodon in the region $\mathcal{R}$:

\begin{equation} \label{EVOL} 
   V_{\mathrm{hodon}}\!\left(\mathcal{R}\right)
   =
   \frac{V\!\left(\mathcal{R}\right)}{\mathrm{E} \left[ N\!\left(\mathcal{R}\right) \right]}
   \,\,{\geq}\!\!\!\!_{\sim}\,\,
   4\ell \ell_{\mathrm{P}}^2
   \;\;\;\;  .
\end{equation}

Assuming each hodon occupies a roughly cubic region, the characteristic linear uncertainty associated with a single hodon is: $\Delta\ell\sim(V_{\mathrm{hodon}}\!\left(\mathcal{R}\right))^{\frac{1}{3}}$. This leads to a spatial uncertainty relation:

\begin{equation} \label{KREL} 
   \Delta\ell
   \,\,{\geq}\!\!\!\!_{\sim}\,\,
   \sqrt[3]{\ell \ell_{\mathrm{P}}^2}
   \;\;\;\;  ,
\end{equation}

which bounds the resolution of length measurements in terms of the region size $\ell$ and the Planck scale $\ell_{\mathrm{P}}$.

This relation is derived from the statistical dispersion of hodon counts and serves as the operational bridge between discrete structure and continuous observables. It is known as the Karolyhazy relation \cite{Karolyhazy}, and has been independently derived by several other authors \cite{Sasakura, Ng, Arzano, Aguilar}.

The Eq.\,(\ref{KREL}) implies that the minimum resolvable length in a region of size $\ell$ is not constant, but scales with $\ell$ itself. It reflects the \textbf{foamy nature of space}: larger regions admit more hodons, but the uncertainty per unit length grows sublinearly. Unlike Heisenberg's uncertainty, this is a geometric uncertainty, arising from the finite density of spatial degrees of freedom.

\subsection{Emergence of continuum geometry}

In the limit $\mathrm{E} \left[ N\!\left(\mathcal{R}\right) \right] \!\!\gg\!\! 1$, the discrete structure becomes densely packed, and classical geometry emerges as a smooth approximation: $\Delta\ell\,\,{\geq}\!\!\!\!_{\sim}\,\,\varepsilon\ell$, where $\varepsilon = \sqrt[3]{\frac{\ell_{\mathrm{P}}^2}{\ell^2}}\to 0$.

Metric quantities such as distance, curvature, and volume are then interpreted as ensemble averages over hodon configurations:

\begin{equation} 
   \mathrm{g}_{\mu\nu}^{\mathrm{eff}(x)}
   =
   \lim_{\mathrm{E} \left[ N\!\left(\mathcal{R}\right) \right]\gg 1}{\langle\mathrm{g}_{\mu\nu}^{\mathrm{hodon}(x)}\rangle}
   \;\;\;\;  ,
\end{equation}

where $\mathrm{g}_{\mu\nu}^{\mathrm{eff}(x)}$ represents the effective metric tensor at a spacetime point $x$, $\mathrm{g}_{\mu\nu}^{\mathrm{hodon}(x)}$ denotes the local metric contribution from hodon-induced spatial fluctuations, and the angle brackets $\langle\:\rangle$ denote an ensemble average. This averaging is meaningful only when the expected number of hodons in the region $\mathcal{R}$ is large. In other words, the effective metric is defined in the thermodynamic limit of many hodons, where statistical fluctuations become negligible and a smooth geometry emerges.

\section{Strong vs. weak holography}

The holographic principle, originally formulated in the context of black hole thermodynamics and later formalized through AdS/CFT duality, asserts that the degrees of freedom within a spatial region are bounded by the area of its boundary, not its volume \cite{Ammon}. In its strongest form, this principle implies a duality between bulk gravitational dynamics and boundary quantum field theory — a correspondence that requires specific spacetime geometries (e.g., asymptotically AdS) and causal structure.

In contrast, the hodon framework adopts a weak holographic stance, using the principle as an operational constraint on the number of distinguishable spatial degrees of freedom in a bounded region. This approach does not invoke duality or boundary field theories but instead treats holography as a counting bound that governs the granularity of space.

\subsection{Strong holography (SH)}

Such holography requires the presence of Lorentzian spacetime with well-defined causal horizons. SH interprets entropy as the logarithm of microstates in a boundary theory. Predictive power of SH derives from exact duality: bulk dynamics are encoded in boundary observables.

\subsection{Weak holography (WH)}

This type of holographic bond applies to all bounded regions, including Euclidean domains. WH interprets entropy as a limit on the number of discrete spatial units (hodons), so its predictive ability comes from statistical limits on measurement resolution.

\section{Integration of Bousso's covariant entropy bound}

To strengthen the physical justification of our holographic constraint, we now incorporate Bousso's covariant entropy bound \cite{Bousso}, which generalizes the holographic principle to arbitrary spacetime regions using light-sheets.

Bousso's bound states that the entropy $S$ passing through any light-sheet $L$ of a surface $B$ satisfies: $S(L)\le\frac{A(B)}{4\ell_{\mathrm{P}}^2}$, where $A(B)$ is the area of the surface generating the light-sheet. This formulation is independent of the spacetime’s global geometry and applies to both static and dynamical settings.

\subsection{Application to hodon framework}

We reinterpret the bound as a constraint on the maximum number of distinguishable spatial imprints (hodons) that can be encoded within a probabilistically causal domain bounded by $B$.

In regions where light-sheets can be constructed (e.g., spherical cavities in Minkowski or weakly curved spacetime), the expected number of hodons is bounded by the area of the generating surface. This allows us to extend the hodon counting principle to {non-Euclidean and dynamical geometries}, preserving consistency with probabilistically causal structure. Hodons are no longer tied to static Euclidean volumes, but to {probabilistically causal accessible domains}, aligning the framework with relativistic principles.

\section{Dark matter}

A hodon mapped into geometrical space can be interpreted as an elementary particle imparted in it. Let us investigate the physical and statistical properties of such a particle.

\subsection{Dimensional anchoring via vacuum energy density}

Assuming that in the region of the observable universe $\mathcal{R}_{\mathrm{U}}$ each hodon contributes a localized energy quantum $E_{\mathrm{hodon}}(\mathcal{R}_{\mathrm{U}})$, we define the hodon mass via $m_{\mathrm{hodon}}(\mathcal{R}_{\mathrm{U}})=E_{\mathrm{hodon}}(\mathcal{R}_{\mathrm{U}})/c^2$. As to the energy quantum $E_{\mathrm{hodon}}(\mathcal{R}_{\mathrm{U}})$, it can be described in terms of the ratio of the observed vacuum energy to the expected count of hodons in $\mathcal{R}_{\mathrm{U}}$:

\begin{equation} 
   E_{\mathrm{hodon}}(\mathcal{R}_{\mathrm{U}})
   =
   \frac{E_{\mathrm{vac}}(\mathcal{R}_{\mathrm{U}})}{\mathrm{E} \left[ N\!\left(\mathcal{R}_{\mathrm{U}}\right) \right]}
   \;\;\;\;  .
\end{equation}

Denote the observed vacuum energy density by $\rho_{\mathrm{vac}}$. From general relativity, it follows that $\rho_{\mathrm{vac}} ={c^4 \Lambda_{\mathrm{vac}} }/{8\pi G}$. Here, the cosmological constant $\Lambda_{\mathrm{vac}}$ is empirically estimated as

\begin{equation} 
   \Lambda_{\mathrm{vac}}
   \sim
   \frac{1}{\ell_{\mathrm{H}}^2}
   \;\;\;\;  ,
\end{equation}

where $\ell_{\mathrm{H}}$ is the Hubble length. Using the Eq.\,(\ref{BOUND}), the expected count of hodons in the region $\mathcal{R}_{\mathrm{U}}$ can be defined as:

\begin{equation} \label{EXPC} 
   \mathrm{E} \left[ N\!\left(\mathcal{R}_{\mathrm{U}}\right) \right]
   \,{\leq}\!\!\!\!_{\sim}\,
   \frac{\ell_{\mathrm{U}}^2}{4\ell_{\mathrm{P}}^2}
   \;\;\;\;  ,
\end{equation}

where $\ell_{\mathrm{U}}$ denotes the characteristic length of the observable universe.

At cosmological scales, the spatial geometry is well approximated by a three-dimensional, flat Riemannian manifold \cite{Liddle}. Accordingly, the region $\mathcal{R}_{\mathrm{U}}$ can be modelled as a three-dimensional Euclidean ball of characteristic length $\ell_{\mathrm{U}} \approx\ell_{\mathrm{H}}$.

Substituting into the energy relation:

\begin{equation} \label{MH} 
   m_{\mathrm{hodon}}(\mathcal{R}_{\mathrm{U}})
   \,\,{\geq}\!\!\!\!_{\sim}\,\,
   \frac{c^2 \Lambda_{\mathrm{vac}}}{8\pi G}
   \cdot
   \ell_{\mathrm{H}}^3
   \cdot
   \frac{4\ell_{\mathrm{P}}^2}{\ell_{\mathrm{H}}^2}
   \sim
   \frac{1}{2\pi}
   \cdot
   \frac{\hbar}{c\ell_{\mathrm{H}}}
   \;\;\;\;  .
\end{equation}

\subsection{Inferring hodon mass from the critical density}

The last relation is consistent with the mass scale inferred from the critical density of the universe.

Using the Friedmann equations, the critical density $\rho_{\mathrm{crit}}$ is defined by: $\rho_{\mathrm{crit}}={3H_0^2}/{8\pi G}$, where $H_0$ is the Hubble constant, expressible as $H_0 =c/\ell_{\mathrm{H}}$. Observational data \cite{Planck} indicate that the dark sector constitutes approximately 96\% of the total energy budget, and in a spatially flat universe, this total closely matches $\rho_{\mathrm{crit}}$. We therefore approximate: $\rho_{\mathrm{DS}}\approx\rho_{\mathrm{crit}}$.

From Eq.\,(\ref{EVOL}), the effective volume per hodon within the observable universe is bounded below by:

\begin{equation} 
   V_{\mathrm{hodon}}\!\left(\mathcal{R}_{\mathrm{U}}\right)
   \,\,{\geq}\!\!\!\!_{\sim}\,\,
   4\ell_{\mathrm{H}}\ell_{\mathrm{P}}^2
   \;\;\;\;  .
\end{equation}

Multiplying the dark sector density by the effective volume per hodon yields:

\begin{equation} 
   m_{\mathrm{hodon}}(\mathcal{R}_{\mathrm{U}})
   =
   \rho_{\mathrm{DS}}
   \cdot
   V_{\mathrm{hodon}}\!\left(\mathcal{R}_{\mathrm{U}}\right)
   \,\,{\geq}\!\!\!\!_{\sim}\,\,
   \frac{3c^2 }{8\pi G\ell_{\mathrm{H}}^2}
   \cdot
   4\ell_{\mathrm{H}}\ell_{\mathrm{P}}^2
   \;\;\;\;  .
\end{equation}

Simplifying, we obtain:

\begin{equation} 
   m_{\mathrm{hodon}}(\mathcal{R}_{\mathrm{U}})
   \,\,{\geq}\!\!\!\!_{\sim}\,\,
   \frac{3}{2\pi}
   \cdot
   \frac{\hbar}{c\ell_{\mathrm{H}}}
   \;\;\;\;  ,
\end{equation}

which is comparable to the mass scale derived from holographic bounds and vacuum energy density. This highlights the inverse dependence on the Hubble radius and situates the hodon mass within a geometric-statistical framework, rather than a particle-based one.

\subsection{Curvature modifies local volume}

In curved spacetime, the effective volume element is given by: $\upd V=\sqrt{-\mathrm{g}}\,\upd^3 x$, where $\mathrm{g}=\det{\mathrm{g}_{\mu\nu}}$ is the determinant of the metric tensor. For a region $\mathcal{R}$, the total volume becomes: $V(\mathcal{R})=\int_{\mathcal{R}}\sqrt{-\mathrm{g}}\,\upd^3 x$. This implies that the volume associated with a single hodon, $V_{\mathrm{hodon}}(\mathcal{R})$, is no longer constant -- it scales with the local curvature encoded in $\mathrm{g}_{\mu\nu}$:

\begin{equation} 
   V_{\mathrm{hodon}}(\mathcal{R})
   =
   \frac{\int_{\mathcal{R}}\sqrt{-\mathrm{g}}\,\upd^3 x}{\mathrm{E} \left[ N\!\left(\mathcal{R}\right) \right]}
   \;\;\;\;  .
\end{equation}

In regions of strong curvature (e.g., near a black hole), $V_{\mathrm{hodon}}(\mathcal{R})$ contracts, increasing the local energy density.

\subsection{Vacuum energy density becomes position-dependent}

In curved spacetime, the vacuum energy density $\rho_{\mathrm{vac}}(x)$ may receive curvature-dependent corrections. For example, in semiclassical gravity: $\rho_{\mathrm{vac}}(x) = \rho_{0}+\alpha \mathrm{R}(x)+\beta \mathrm{R}_{\mu\nu}(x)\mathrm{R}^{\mu\nu}(x)+\cdots$, where $\mathrm{R}(x)$ is the Ricci scalar and $\alpha$, $\beta$ are model-dependent coefficients. This means the energy contribution of a hodon becomes:

\begin{equation} 
   E_{\mathrm{hodon}}(x)
   =
   \rho_{\mathrm{vac}}(x)
   \cdot
   V_{\mathrm{hodon}}(x)
   \;\;\;\;  ,
\end{equation}

and the corresponding mass scales as: $m_{\mathrm{hodon}}(x)={\rho_{\mathrm{vac}}(x)}⁄{c^2}\cdot V_{\mathrm{hodon}}(x)$.

\subsection{Scaling behavior near strong curvature}

Near gravitational wells (e.g., black holes), curvature increases: $V_{\mathrm{hodon}}(x)$ decreases due to spatial contraction, while $\rho_{\mathrm{vac}}(x)$ may increase due to curvature-induced vacuum polarization.

Thus, the hodon mass within a region $\mathcal{R}$ of characteristic length $\ell$ scales as:

\begin{equation} 
   m_{\mathrm{hodon}}(x)
   \propto
   \frac{\rho_{\mathrm{vac}}(x)}{c^2}
   \cdot
   \sqrt{-\mathrm{g}(x)}
   \cdot
   \ell\ell_{\mathrm{P}}^2
   \;\;\;\;  .
\end{equation}

This scaling reflects the dual dependence on local geometry and quantum vacuum structure. In flat spacetime, this reduces to the original estimate: $m_{\mathrm{hodon}}(\mathcal{R})\sim \rho_{\mathrm{vac}} ⁄c^2 \cdot \ell\ell_{\mathrm{P}}^2$.

Therefore, the hodon mass is not a fixed scalar -- it’s a local functional of spacetime curvature. In regions of high curvature, hodons become more energetically dense, potentially contributing to gravitational backreaction. This scaling supports the idea that dark matter effects may emerge from curvature-dependent hodon statistics, without invoking new particles.

\subsection{Stability of hodon mass}

Returning to the Eq.\,(\ref{MH}). Using numerical values, we obtain that $m_{\mathrm{hodon}}(\mathcal{R}_{\mathrm{U}})$ is extraordinarily small: its greatest lower bound is about $10^{-65} \un{g}$ ($\sim 10^{-32}\un{eV}$). If hodons comprise the dark sector of the observable universe, then the observed discrepancy between baryonic and gravitational mass must arise from their overwhelming abundance. And indeed, according to Eq.\,(\ref{EXPC}), the maximum expected number of hodons in the region $\mathcal{R}_{\mathrm{U}}$ yields $\sim 10^{123}$ -- a staggering number.

The infimum of $m_{\mathrm{hodon}}(\mathcal{R}_{\mathrm{U}})$ is many orders of magnitude below the mass of any known particle, including ultralight dark matter candidates (e.g., fuzzy DM with $m\sim 10^{-22}\,\un{eV}$ \cite{Ferreira}). Consequently, a hodon is intrinsically stable: there exists no lighter particle into which it could decay. It remains in its original quantum state, immune to energy loss through decay.

\subsection{Non-relativity of hodons}

In standard relativistic kinematics, a particle with mass $m$ and momentum $p$ has total energy: $E=\sqrt{(pc)^2+(mc^2)^2}$. For hodons, any finite momentum $p$ would immediately dominate the rest energy term $mc^2$, implying relativistic behavior. However, no such momentum is defined: hodons are not excitations of a field, nor do they possess translational degrees of freedom. Their mass is not associated with motion, but with spatial embedding. Thus, hodons cannot be classified as relativistic particles. They are ultra-light, but not ultra-fast.

In quantum mechanics, translational motion implies a wavefunction $\psi(x,t)$ with non-zero group velocity: $v_g =d\omega⁄dk$. For a free particle, this corresponds to kinetic energy $T=\frac{p^2}{2m}$. But hodons are not described by propagating wavefunctions. They are spatial units -- non-material, non-localized, and non-dynamical. Their distribution is governed by statistical geometry, not by Hamiltonian evolution. Moreover, the hodon framework lacks a dispersion relation. There is no defined $\omega(k)$, no phase velocity, and no group velocity. Consequently, hodons do not propagate, oscillate, or translate. They are effectively stationary.

The energy associated with a hodon, $E_{\mathrm{hodon}}(\mathcal{R}_{\mathrm{U}})=\rho_{\mathrm{vac}}(\mathcal{R}_{\mathrm{U}})\cdot V_{\mathrm{hodon}}(\mathcal{R}_{\mathrm{U}})$,  arises from its embedding in the vacuum structure. This energy is not kinetic, thermal, or field-based. It is geometric -- an emergent consequence of spacetime discreteness and holographic entropy bounds. Since hodons do not move, their energy cannot be partitioned into kinetic and potential components. There is no dynamical evolution, no acceleration, and no inertial frame in which a hodon has velocity. Thus, hodons are kinematically non-relativistic not because they move slowly, but because they do not move at all. Their inertness is ontological, not dynamical.

In summary, although hodons possess an ultra-light effective mass, they are not free particles. They lack translational degrees of freedom, do not obey relativistic dispersion relations, and carry no kinetic energy. Their energy content is embedded in vacuum fluctuations and spatial uncertainty. Consequently, hodons are kinematically non-relativistic: effectively stationary and inert.

\subsection{A minimal toy model}

To analyze their gravitational behavior, consider a stable system in the observable universe composed of an elementary particle (e.g., a photon or proton) surrounded by $n$ hodons, bound by a conservative force $F$. Let the elementary particle be fixed at the origin.

According to the non-relativistic virial theorem \cite{Goldstein, Sivardiere}, the time-averaged total kinetic energy of the system is given by:

\begin{equation} 
   \langle T \rangle
   =
   -
   \frac{1}{2}
   \cdot
   \Big\langle \sum_{k=1}^{n} \vect{F}_k\cdot\vect{r}_k \Big\rangle
   \;\;\;\;  .
\end{equation}

Here, $\vect{F}_k$ is the net force acting on the $k$th hodon, located at position $\vect{r}_k$, and the angle brackets $\langle\:\rangle$ denote the time average of the enclosed quantity. The force $\vect{F}_k$ is the sum:\smallskip

\begin{equation} 
   \vect{F}_k
   =
   \vect{F}_{0k}
   +
   \sum_{j=1}^{n} \vect{F}_{jk}
   \;\;\;\;  ,
\end{equation}

where $\vect{F}_{0k}$ is the force from the elementary particle, and $\vect{F}_{jk}$ is the force from hodon $j\in\{1\dots,n\}$. Thus, total kinetic energy separates into two contributions:

\begin{equation} 
   \langle T \rangle
   =
   \langle T_{\mathrm{particle-hodon}} \rangle
   +
   \langle T_{\mathrm{hodon-hodon}} \rangle
   \;\;\;\;  .
\end{equation}

The first term concerns the total kinetic energy of particle-hodon interactions:

\begin{equation} 
   \langle T_{\mathrm{particle-hodon}} \rangle
   =
   -
   \frac{1}{2}
   \cdot
   \Big\langle \sum_{k=1}^{n} \vect{F}_{0k}\cdot\vect{r}_{k0} \Big\rangle
   \;\;\;\;  ,
\end{equation}

where $\vect{r}_{k0}$ denotes the position of the $k$th hodon with respect to the position of the elementary particle.

On the assumption that hodons do not act on themselves (i.e., $\vect{F}_{jj}=0$), the term for the total kinetic energy of hodon-hodon interactions can be expressed as:

\begin{equation} 
   \langle T_{\mathrm{hodon-hodon}} \rangle
   =
   -
   \frac{1}{2}
   \cdot
   \Big\langle \sum_{k=2}^{n} \sum_{j=1}^{k-1} \vect{F}_{jk}\cdot\vect{r}_{kj} \Big\rangle
   \;\;\;\;  ,
\end{equation}

where $\vect{r}_{kj} = \vect{r}_{k} - \vect{r}_{j}$ represents the relative position of hodon $k$ with respect to hodon $j$. Newton's third law of motion, i.e., $\vect{F}_{jk} =-\vect{F}_{kj}$, has been applied to simplify the expression. 

Assuming stationarity and sufficient time for averaging, we replace the time average  $\langle T \rangle$ with the expectation $\mathrm{E} \left[ T \right]$. Specifically, let the random variable $N(\vect{r})$ denote the number of hodons embedded at a given position $\vect{r} \!\in\! \{\vect{r}_{k0},\vect{r}_{kj}\}$. Then, the total kinetic energy can be expressed as:

\begin{equation} 
   \langle T \rangle
   \approx
   \mathrm{E} \left[ T \right]
   =
   -
   \frac{1}{2}
   \cdot
      \sum \vect{F}\cdot\vect{r}
      \cdot P \!\left\{ N(\vect{r}) = 1 \right\}
   \;\;\;\;  .
\end{equation}

Here, $P \!\left\{ N(\vect{r}) = 1 \right\}$ is the probability of finding exactly one hodon at position $\vect{r}$. According to Eq.\,(\ref{SPRIN}), this probability is given by:

\begin{equation} 
  P \!\left\{ N(\vect{r}) = 1 \right\}
   =
   \Lambda(\vect{r}) \cdot e^{-\Lambda(\vect{r})}
   \;\;\;\;  ,
\end{equation}

where $\Lambda(\vect{r})$ denotes the intensity parameter of the Poisson point process. In a geometrical space that is homogeneous and isotropic on cosmological scales, the density of embedded hodons must be spatially uniform. Consequently, the probability of finding a hodon at a given position $\vect{r}$ depends only on the magnitude $r = \left| \vect{r} \right|$ -- the radial distance from a reference point -- and not on the direction. This implies that the intensity parameter $\Lambda(\vect{r})$ is proportional to $V(r)$, the volume of a ball of radius $r$. Hence, the detection probability simplifies to:

\begin{equation} 
   P \!\left\{ N(\vect{r}) = 1 \right\}
   =
   \varrho_V
   \cdot
   V\!\left(r\right)
   \cdot
   e^{-\varrho_V \cdot V\!\left(r\right)}
   \,\propto\:
   r^3
   \cdot
   e^{-r^3}
   \;\;\;\;  .
\end{equation}

Given that hodons possess only two measurable attributes -- information and mass-energy -- they interact with photon-baryon matter or other hodons exclusively via gravity. Accordingly, the attractive force $\vect{F}\in\{\vect{F}_{0k},\vect{F}_{jk}\}$ is assumed to act in the opposite direction to the position vector $\vect{r}$ and to vary as $-\frac{1}{r^2}$. In consequence, the virial can be written as:

\begin{equation} \label{TKE} 
   \langle T \rangle
   \propto
   \frac{1}{2}
   \cdot
   \sum_{k=1}^{n} r_{k0}^2 \cdot e^{-r_{k0}^3}
   +
   \frac{1}{2}
   \cdot
   \sum_{k=2}^{n} \sum_{j=1}^{k-1} r_{kj}^2 \cdot e^{-r_{kj}^3}
   \;\;\;\;  .
\end{equation}

Here, each term describes a probabilistically weighted interaction across spatial separation $r$. 

At short distances ($r \to 0$), contributions vanish polynomially due to the $r^2$ prefactor. Consequently, hodons are effectively incapable of colliding with baryonic particles or one another, as such interactions require vanishing separations. This dynamical exclusion precludes the formation of tightly bound configurations -- such as stars, planetary systems, or black holes -- from hodons alone. 

Furthermore, Eq.\,(\ref{TKE}) implies that hodons do not accumulate in central galactic regions, where the baryonic core coincides with the gravitational potential minimum. The lack of spectral enhancement near $r \to 0$ aligns with observational data indicating that dark matter halos exhibit \textbf{cored} profiles, rather than \textbf{cusped} ones -- as noted by \cite{Moore, Bosch, Blok}, and others. 

Conversely, at the opposite extreme ($r \gg 1$), the exponential term $e^{-r^3}$ suppresses long-range contributions, preventing indefinite dispersion. This balance -- between short-range polynomial damping and long-range exponential decay -- promotes gravitational clumpiness without collapse, favouring extended, low-density structures. Such behavior mirrors the large-scale morphology of dark matter halos: diffuse, stable, and spatially extended \cite{Klypin}.

\subsection{Halo core profiles and rotation curves}

The statistical nature of hodon-induced clumpiness leads to smooth halo cores rather than cuspy central densities. Since hodons do not interact or self-annihilate, their distribution remains diffuse, avoiding the over-concentration typical of cold dark matter simulations. This feature aligns with observations of flat galactic rotation curves and core-dominated mass profiles in dwarf galaxies. The absence of particle dynamics in the hodon framework provides a clean resolution to the cusp-core problem without requiring baryonic feedback or exotic interactions.

\subsection{Void stability and cosmic web morphology}

In regions of low hodon density, gravitational collapse is statistically disfavoured. This leads to enhanced stability of cosmic voids, which resist fragmentation and maintain coherent underdensity profiles. The framework predicts a distinct void size distribution and morphology, potentially observable in large-scale galaxy surveys. Moreover, the granular structure of space may influence the filamentary architecture of the cosmic web, introducing subtle deviations from ${\Lambda}$CDM predictions in connectivity and anisotropy \cite{Cintio}.

\section{Covariant hodon density and entropic evolution}

To address the absence of formal dynamics for hodons, we now introduce a covariant statistical representation of hodon density and propose an entropy-driven evolution equation consistent with general relativity. This formulation preserves the ontological stance of the hodon framework -- namely, that hodons are discrete spatial degrees of freedom rather than particles -- while embedding their behaviour within a relativistically consistent structure.

\subsection{Covariant hodon density field}

Let $\mathcal{N}(x^\mu)$ be a scalar field defined over spacetime, representing the local density of hodons per unit volume. In the static limit, $\mathcal{N}(x^\mu)$ can be identified as the local sprinkling density $\varrho(x)$, where $x$ is treated as a spacetime point and $\mathcal{N}(x^\mu)$ is evaluated on a spacelike hypersurface. This identification holds when the hodon density is treated as a passive statistical field, not a dynamical entity. In the general case, however, $\mathcal{N}(x^\mu)$ is a field-theoretic upgrade of $\varrho(x)$: it allows for curvature coupling, entropy production, and causal evolution. 

Since hodons are non-material and non-dynamical, $\mathcal{N}(x^\mu)$ does not obey a conventional continuity equation. Instead, it encodes the statistical availability of spatial degrees of freedom at each point $x^\mu$, constrained by holographic bounds and curvature.

We define the effective stress-energy tensor associated with hodon fluctuations as: 

\begin{equation} 
   \tens{T}_{\mathrm{hodon}}^{\mu\nu}
   =
   \alpha\mathcal{N}(x^\mu)\mathrm{g}^{\mu\nu}
   +
   \beta\nabla^{\mu}\mathcal{N}\nabla^{\nu}\mathcal{N}
   \;\;\;\;  ,
\end{equation}

where $\alpha$ and $\beta$ are constants determined by the scaling of vacuum energy and spatial uncertainty, and $\mathrm{g}^{\mu\nu}$ is the spacetime metric. This tensor contributes to Einstein's field equations: 

\begin{equation} 
   \tens{G}^{\mu\nu}
   =
   8\pi G 
   \left( 
      \tens{T}_{\mathrm{matter}}^{\mu\nu}
      +
      \tens{T}_{\mathrm{hodon}}^{\mu\nu} 
   \right)
   \;\;\;\;  ,
\end{equation}

allowing hodons to influence curvature without invoking particle-like dynamics.

\subsection{Entropy-based evolution equation}

We now propose an evolution equation for $\mathcal{N}(x^\mu)$ based on entropy gradients. Let $S(x^\mu)$ be the local entropy density associated with spatial uncertainty, defined via:

\begin{equation} \label{LED} 
   S(x^\mu)
   =
   \gamma\cdot\mathcal{N}(x^\mu)
   \;\;\;\;  ,
\end{equation}

where $\gamma$ is a scaling factor derived from the holographic entropy per hodon. Assuming entropy flows along causal trajectories, we posit the following covariant evolution equation:

\begin{equation} 
   \nabla_{\mu}\left( S(x^\mu) u^{\mu} \right)
   =
   \sigma(x^\mu)
   \;\;\;\;  ,
\end{equation}

where $u^{\mu}$ is the local four-velocity of the observer field, and $\sigma(x^\mu)$ is a source term representing entropy production due to curvature or matter interactions. Substituting the Eq.\,(\ref{LED}), we obtain:

\begin{equation} 
   \nabla_{\mu}
   \left( 
      \mathcal{N}(x^\mu) u^{\mu} 
   \right)
   =
   \frac{\sigma(x^\mu)}{\gamma}
   \;\;\;\;  .
\end{equation}

This equation governs the evolution of hodon density in response to entropy flux, allowing structure formation to be modelled as a statistical process driven by geometric constraints.

In flat spacetime with no entropy production, $\sigma(x^\mu)=0$, and hodon density remains constant along flow lines. In curved regions, $\sigma(x^\mu)>0$, leading to local accumulation of hodons and gravitational clumpiness.

\subsection{Interpretation and future development}

This formulation provides a foundation for embedding hodon dynamics into relativistic cosmology. It allows:

\begin{itemize}
   \item Hodon density to evolve in response to curvature and entropy gradients.
   \item Contributions to gravitational field equations via $\tens{T}_{\mathrm{hodon}}^{\mu\nu}$.
   \item Statistical structure growth without invoking particle trajectories.
\end{itemize}

Future work will explore whether this evolution equation admits attractor solutions, supports halo formation, and reproduces observed large-scale structure when coupled to baryonic matter and radiation.

\section{Final thoughts: Empirical and philosophical synthesis}

The hodon framework offers a radical yet empirically grounded reinterpretation of the dark sector. By treating spacetime as composed of discrete, non-material units whose statistical distribution governs gravitational behavior, the model circumvents the need for new particles or modified gravity. Through virial-based modeling, entropy-driven evolution, and holographic constraints, it reproduces key features of dark matter phenomenology -- cored halo profiles, suppression of small-scale structure, and void stability -- while remaining consistent with relativistic principles and observational bounds. Future work will focus on embedding these mechanisms into simulation pipelines and refining their predictive scope across cosmological observables.

\end{document}